\pdfoutput=1

\documentclass[11pt]{article}

\usepackage{emnlp2021}

\usepackage{times}
\usepackage{latexsym}
\usepackage{graphicx}
\usepackage{amsmath}
\usepackage{hyperref}
\DeclareMathOperator*{\argmax}{argmax}

\usepackage[T1]{fontenc}

\usepackage[utf8]{inputenc}

\usepackage{microtype}
\usepackage{amsfonts}

\title{A Unified Speaker Adaptation Approach for {ASR}}

\author{Yingzhu Zhao$^{1, 2}$\thanks{$^*$Yingzhu Zhao is under the Joint PhD Program between Alibaba and Nanyang Technological University.} \ , Chongjia Ni$^2$ , Cheung-Chi Leung$^2$ \\ {\bf Shafiq Joty$^1$, Eng Siong Chng$^1$ , Bin Ma$^2$}  \\
  $^1$Nanyang Technological University, Singapore \\
  $^2$Machine Intelligence Technology, Alibaba Group \\
  \{srjoty, aseschng\}@ntu.edu.sg    \\   \{yingzhu.zhao, ni.chongjia, cc.leung, b.ma\}@alibaba-inc.com       \\}

\begin{document}
\maketitle
\begin{abstract}
Transformer models have been used in automatic speech recognition (ASR) successfully and yields state-of-the-art results. However, its performance is still affected by speaker mismatch between training and test data. Further finetuning a trained model with target speaker data is the most natural approach for adaptation, but it takes a lot of compute and may cause \emph{catastrophic forgetting} to the existing speakers. In this work, we propose a unified speaker adaptation approach consisting of feature adaptation and model adaptation. For feature adaptation, we employ a speaker-aware persistent memory model which generalizes better to unseen test speakers by making use of speaker i-vectors to form a persistent memory. For model adaptation, we use a novel gradual pruning method to adapt to target speakers without changing the model architecture, which to the best of our knowledge, has never been explored in ASR. Specifically, we gradually prune less contributing parameters on model encoder to a certain sparsity level, and use the pruned parameters for adaptation, while freezing the unpruned parameters to keep the original model performance. We conduct experiments on the Librispeech dataset. Our proposed approach brings relative 2.74-6.52\% word error rate (WER) reduction on general speaker adaptation. On target speaker adaptation, our method outperforms the baseline with up to 20.58\% relative WER reduction, and surpasses the finetuning method by up to relative 2.54\%. Besides, with extremely low-resource adaptation data (e.g., 1 utterance), our method could improve the WER by relative 6.53\% with only a few epochs of training.
\end{abstract}

\section{Introduction}
\label{sec:intro}

End-to-end models yield state-of-the-art performance on automatic speech recognition (ASR)  in the past decade, such as connectionist temporal classification (CTC) model \citep{literature1,literature3}, attention-based encoder-decoder model \citep{literature2}, recurrent neural network transducer (RNN-T) \citep{literature3}, transformer model \citep{Speech-transformer} and conformer model \citep{conformer}. However, model performance deteriorates due to \emph{speaker mismatch} between training and test data. Given the target speaker, finetuning the trained model could alleviate the speaker mismatch problem to some extent, but finetuning the entire model requires large amounts of compute to be effective, and it could in turn bring catastrophic forgetting \cite{catastrophic} to the existing speakers.

Currently, there are two lines of studies to address the speaker mismatch problem in neural network based models. One category is working on the acoustic features, i.e., either by normalizing acoustic features to be speaker-independent \citep{feature2,feature3,feature4} or by introducing additional speaker related knowledge (e.g., i-vector) to adapt the acoustic model \citep{feature5,feature7,feature8,fanzhiyun}. A summary vector of each utterance can be trained to replace speaker i-vector \citep{feature10}. To adapt to acoustic variability, \citet{feature11} add shifting and scaling parameters in the layer-normalization layer.

The other category belongs to model adaptation, i.e., to train the speaker-dependent model from speaker-independent model parameters with extra adaptation data. To avoid overfitting, techniques such as L2 regularization \citep{l2}, Kullback-Leibler divergence \citep{kld} and adversarial multitask learning \citep{mtl} have been used. Because finetuning the entire model is computationally expensive, \citet{model1,model2,model3} only adapt specific layers or a subset of parameters. In particular, \citet{lhuc1,lhuc2,lhuc3} reparameterize each hidden unit with speaker-dependent amplitude function in fully-connected or convolutional neural network layers. However, it is difficult to determine which model parameters to adapt for target speaker, and choosing certain sub-layer(s) intuitively may not be optimal.

In this work, we propose a unified speaker adaptation model by making use of both feature adaptation and model adaptation. For feature adaptation, we propose the speaker-aware persistent memory model to generalize better to unseen test speakers. In particular, speaker i-vectors from the training data are sampled and concatenated to speech utterances in each encoder layer, and the speaker knowledge is learnt through attention computation with speaker i-vectors. Our method learns utterance level speaker knowledge, which is more effective than learning time step dependent speaker knowledge \citep{fanzhiyun} since it is more robust to various variability factors along an utterance. 

For model adaptation, we explore gradual pruning \citep{prune2}, which to the best of our knowledge, is the first time being studied for speaker adaptation. We gradually prune less contributing parameters on model encoder, and then use the pruned parameters for target speaker adaptation while freeze the unpruned parameters to retain the model performance on general speaker data. In this way, our model could adapt to target speakers very fast by updating only a small percentage (10\%) of encoder parameters, and it does not change the model architecture. Freezing unpruned parameters alleviates the catastrophic forgetting problem as well. 

Our proposed approach brings relative 2.74-6.52\% WER reduction on general speaker adaptation. On target speaker adaptation, our method outperforms the baseline with up to 20.58\% relative WER reduction, and surpasses the finetuning method by up to relative 2.54\%\footnote{Our code is available at \url{https://github.com/zyzpower/gradprune_speaker}}.

\section{Background}

\subsection{Speech Transformer}

Speech transformer \citep{Speech-transformer} is an extension of the transformer model \citep{transformer} for ASR. We briefly introduce the speech transformer model here. For a speech input sequence, speech transformer first applies two convolution layers with stride two to reduce hidden representation length. A sinusoidal positional encoding is added to encode position information. Both encoder and decoder in speech transformer model use multi-head attention network. Attention network has three inputs key, query and value, which are distinct transformations of an input sequence. The multi-head attention network is computed by concatenating single attention network $h$ times:
\begin{equation}
  Attention(Q,K,V) = softmax(\frac{QK^T}{\sqrt{d_k}})V
  \label{attention}
\end{equation}
\begin{equation}
  MultiHd(Q,K,V)=Concat(hd_1,...,hd_h)W^O
  \label{multiheadattention}
\end{equation}
\begin{equation}
  hd_i=Attention(QW_i^Q,KW_i^K,VW_i^V)
  \label{singlehead}
\end{equation}
where $h$ is the head number, $W_i^Q\in\mathbb{R}^{d_{model}\times d_q}$, $W_i^K\in\mathbb{R}^{d_{model}\times d_k}$, $W_i^V\in\mathbb{R}^{d_{model}\times d_v}$, $W^O\in\mathbb{R}^{hd_v\times d_{model}}$, we set $d_k=d_q=d_v=d_{model}/h$.

Multi-head attention could learn input representation in different subspaces simultaneously. For encoder, the three inputs all come from the speech input, so the attention network is called self-attention network. For decoder, the text input first goes through self-attention network. To maintain autoregression in decoder, a mask is applied to future tokens. To incorporate information from the speech input, in the next attention network, key and value vectors come from encoder, and query vector comes from decoder, so this attention network is called cross-attention network. Layer normalization and residual connection are applied before and after multi-head attention network. Afterwards, there is a position-wise feedforward network with rectified linear unit (ReLU) activation:
\begin{equation}
  FFN(x)=max(0,xW_1+b_1)W_2+b_2
  \label{ff}
\end{equation}
where $W_1\in\mathbb{R}^{d_{model}\times d_{ff}}$, $W_2\in\mathbb{R}^{d_{ff}\times d_{model}}$, and the biases $b_1\in\mathbb{R}^{d_{ff}}$, $b_2\in\mathbb{R}^{d_{model}}$. 

Self-attention network and position-wise feedforward network form an encoder layer. There is an additional cross-attention network in a decoder layer. There are $N_e$ encoder layers and $N_d$ decoder layers in total.

\subsection{Speaker Adaptation}
\label{sec:speaker_adaptation}

Speaker adaptation arises due to \emph{speaker mismatch} between training and test data. It aims to adapt the model to a target speaker, which is a critical component in HMM-based models \citep{hmm1,hmm2,hmm3,hmm4}. For neural network based models, many approaches are developed as well as discussed briefly in Section~\ref{sec:intro}. 

Adapting an ASR model is a challenging task given that ASR model is a huge and complex model with a large number of parameters to update. Finetuning the entire model takes significant computational resources to reach the optimal performance, and it potentially causes catastrophic forgetting problem \citep{,catastrophic,cf}, which means when model parameters trained for existing speakers are adapted for target speaker, knowledge learnt previously is lost.

\subsection{I-vector}

I-vector is a low-dimension vector that is dependent on speaker and channel. I-vector dimension is fixed as defined no matter how long the utterance is. It is extracted based on a data-driven approach by mapping frames of an utterance to a low-dimensional vector space using a factor analysis technique \citep{ivec1}. The system is based on Support Vector Machines or directly using the cosine distance value as a final decision score \citep{ivec2}. I-vector was initially invented for audio classification and identification, but recently it is used for speaker adaptation as well \citep{feature5,ivec2,ivec3}.

\section{Proposed Method}

We propose an efficient speaker adaptation model by making use of both feature adaptation and model adaptation. For feature adaptation, we embed speaker knowledge represented by a number of fixed speaker i-vectors into each input utterance \citep{zyz1}. This aims to capture speaker information through attention computation between each utterance and speaker i-vectors. For model adaptation, an effective method is employed that can adapt to target speaker very fast without sacrificing performance on existing speakers. In particular, we prune the model gradually and finetune a small subset of parameters to be speaker-specific.

\subsection{Speaker-Aware Persistent Memory for Feature Adaptation}
\label{sec:speaker_aware}

Speaker-aware persistent memory model learns speaker knowledge from i-vectors. We first randomly sample $N$ speaker i-vectors $m_1,...,m_N\in\mathbb{R}^{d_{k}}$ which form the speaker space \cite{speakerspace1,speakerspace2}. Here we have the assumption that the linear combinations of speaker space are enough to cover the speaker information space, i.\,e., any unknown speaker not seen in the training data can be represented approximately by the sampled i-vectors from the training data. We name the learned transformation of speaker space as persistent memory vectors $M_k$ and $M_v$:
\begin{equation}
  M_k=Concat([U_km_1,...,U_km_N])\in\mathbb{R}^{N\times d_k}
  \label{mkey}
\end{equation}
\begin{equation}
  M_v=Concat([U_vm_1,...,U_vm_N])\in\mathbb{R}^{N\times d_k}
  \label{mvalue}
\end{equation}
where $U_k\in\mathbb{R}^{d_k\times d_k}$, $U_v\in\mathbb{R}^{d_k\times d_k}$. Only $U_k$ and $U_v$ matrices are learnable while sampled i-vectors are fixed in this method.

\begin{figure}[t]
  \centering
  \includegraphics[width=\linewidth]{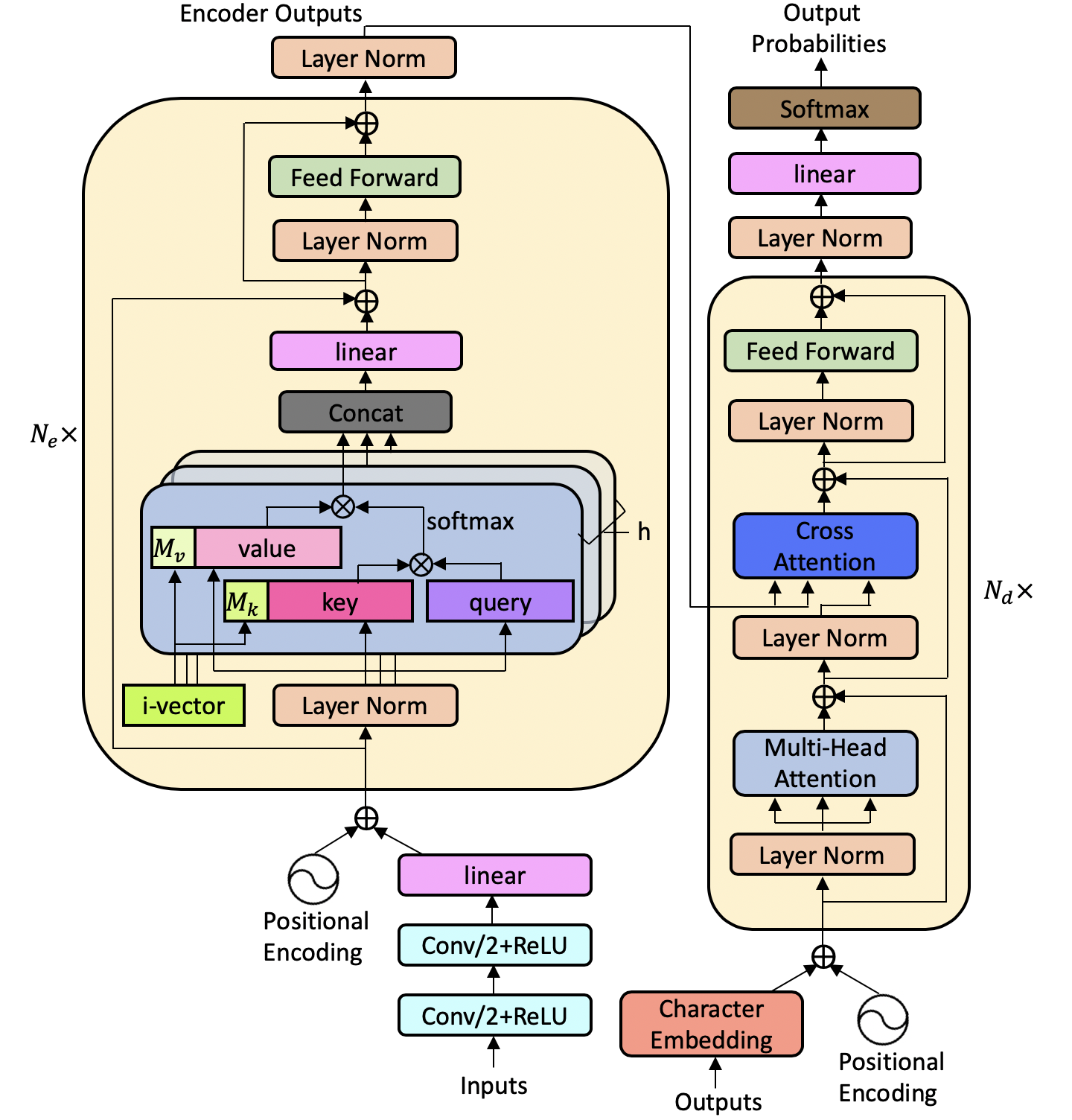}
  \caption{Speaker-aware persistent memory model. $M_k$ and $M_v$ from speaker i-vectors are concatenated to key and value vectors.}
  \label{fig:sepm}
\end{figure}

With the persistent memory vectors, we concatenate them respectively to the input vectors of self-attention network $X=[x_1,...,x_t]$ to be the new key and value vectors. Attention network thus captures speaker-specific knowledge through attention computation between each utterance and persistent memory vectors as Eq.~\ref{all}:
\begin{equation}
  K_m=[k_1,...,k_{t+N}]=([W_kx_1,...,W_kx_t],M_k)
  \label{pmkey}
\end{equation}
\begin{equation}
  V_m=[v_1,...,v_{t+N}]=([W_vx_1,...,W_vx_t],M_v)
  \label{pmvalue}
\end{equation}
\begin{equation}
  Attention(Q,K_m,V_m) = softmax(\frac{QK_m^T}{\sqrt{d_k}})V_m
  \label{all}
\end{equation}

Since $M_k$ and $M_v$ are shared across all layers, they form the persistent memory. Given that persistent memory is meant for capturing speaker knowledge, we name it as speaker-aware persistent memory. The overall framework of speaker-aware persistent memory model is shown in Figure~\ref{fig:sepm}.

Since our method aims at learning any speaker information from the speaker space, it effectively addresses the problem of having unseen speakers in the test data. Furthermore, using static i-vectors saves the effort to compute the i-vectors of all speakers in the training data. Besides, the attention computation (Eq.~\ref{all}) with persistent memory vectors is taken over the entire utterance $x_1$ to $x_t$, so each input speech time step takes part in extracting speaker information. This holistic consideration is more effective than \citet{fanzhiyun} who compute time step dependent speaker representations, which may be more susceptible to various variability factors along an utterance such as speaking rhythm.

\subsection{Gradual Pruning for Model Adaptation}
\label{sec:gradual_pruning}

The speaker-aware persistent memory method discussed above is for general speaker adaptation without knowing target speaker profile. If the target speaker data is available, finetuning the trained model with target speaker data could result in catastrophic forgetting problem as detailed in Section~\ref{sec:speaker_adaptation}. To address this, we take advantage of an effective approach to adapt to target speaker in a fast manner, and retain the model performance on the general speaker data at the same time.

Given an input speech $y=\{y_1,...,y_n\}$ and output text $z=\{z_1,...,z_m\}$, ASR models the conditional probability of the output text over the input speech as follows:
\begin{equation}
  P(z|y;\theta) = \prod_{i=1}^{m}P(z_i|y,z_{<i};\theta)
  \label{asr1}
\end{equation}
where $\theta$ represents the model parameters. Given a training dataset $D=\{y_D^j,z_D^j\}_{j=1}^L$, $\theta$ is trained to maximize the following log-likelihood objective:
\begin{equation}
  \mathcal{J}(\theta)=\argmax_{\theta}\sum_{j=1}^{L}\log P(z_D^j|y_D^j;\theta)
  \label{asr2}
\end{equation}

Given the target speaker dataset $D_t=\{y_{D_t}^j,z_{D_t}^j\}_{j=1}^{L_t}$, directly finetuning the trained model means continuing to train the model to maximize the log-likelihood:
\begin{equation}
  \mathcal{J}(\theta_{D_t})=\argmax_{\theta_{D_t}}\sum_{j=1}^{L_t}\log P(z_{D_t}^j|y_{D_t}^j;\theta_{D_t})
  \label{asr3}
\end{equation}
where $\theta_{D_t}$ is initialized with the trained parameters $\theta$ in Eq.~\ref{asr2}.

\begin{figure}[t]
  \centering
  \includegraphics[width=\linewidth]{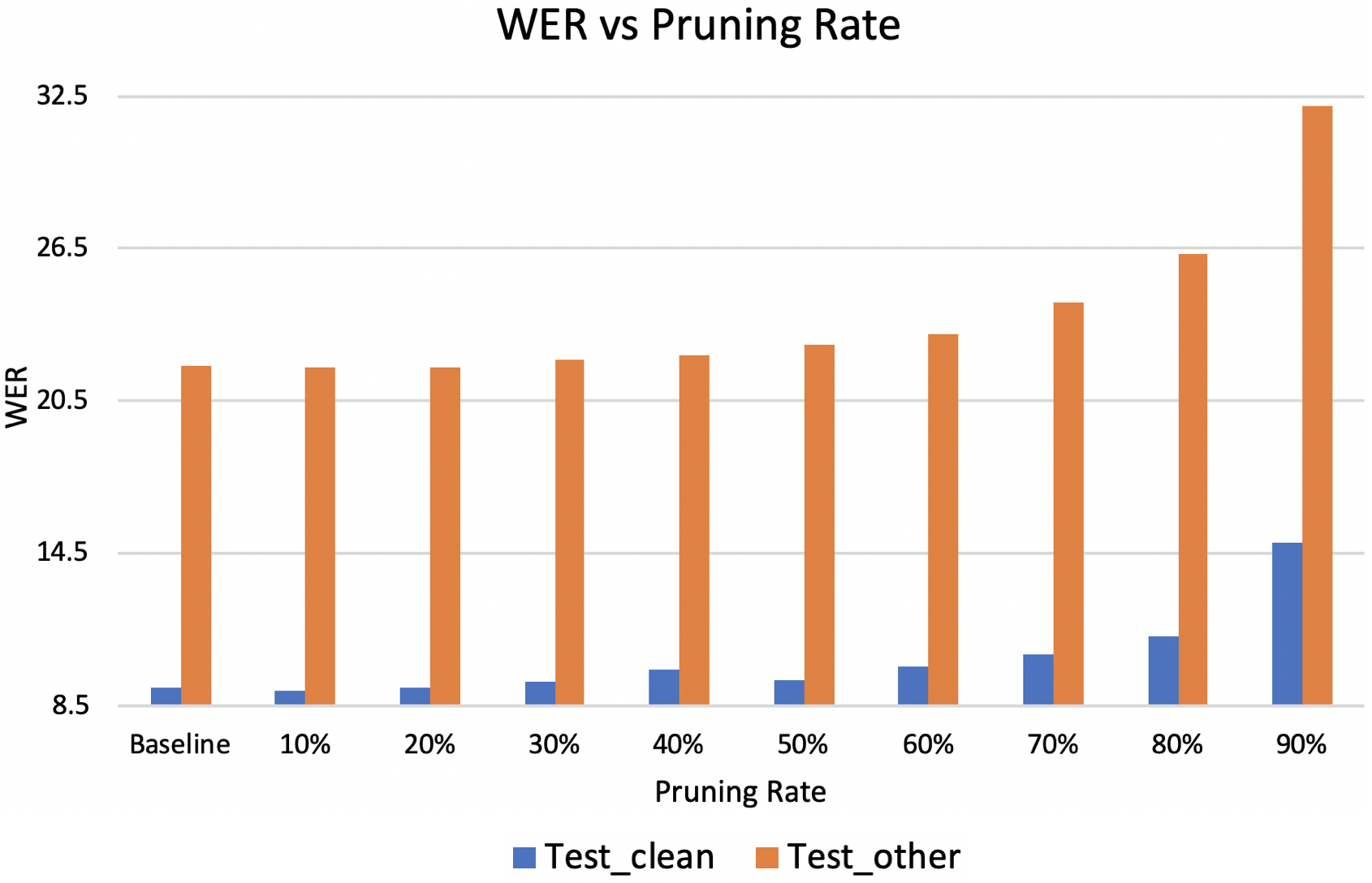}
  \caption{WER results on Librispeech test data with different pruning rates.}
  \label{fig:pr0}
\end{figure}

As shown in many recent studies \citep{prune1,prune2,prune3}, not all parameters in a neural network model contribute to the training objective. Pruning the redundant parameters leads to negligible performance degradation \citep{prune4,prune5} or may even outperform the original model \citep{prune2} due to better generalization. 

Our experiments in Figure~\ref{fig:pr0} show that up to 50\% of encoder parameters can be pruned in ASR with negligible performance degradation. Therefore, we first prune the model with training data gradually to the predetermined sparsity level by zeroing out low magnitude parameters for every 10k training steps, i.e., only retain a certain percentage of high magnitude unpruned parameters $\theta_{UP}$ ultimately. This is to unearth the sub-network whose performance well matches the original model.

\begin{figure*}[t]
  \centering
  \includegraphics[width=\linewidth]{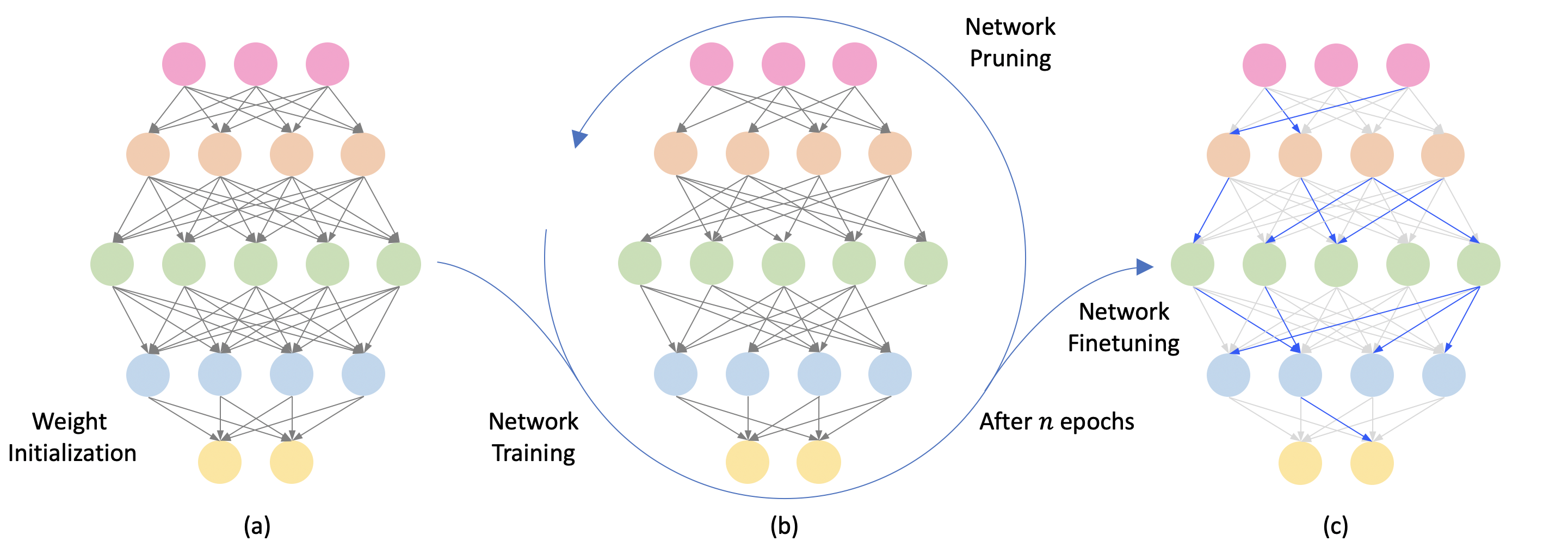}
  \caption{Illustration of gradual pruning method for speaker adaptation. After network parameters are initialized (a), we train and prune the model at the same time for $n$ epochs (b), and lastly finetune the pruned parameters with target speaker data (c). Light gray connections in (c) mean corresponding parameters are frozen, while blue ones indicate parameters are finetuned, which are the parameters pruned at earlier stage (b).}
  \label{fig:prune}
\end{figure*}

Different from \citet{prune-tune} who prune on a well-trained model, we train and prune concurrently as seen in Figure~\ref{fig:prune}(b) (with warm-up training first) to reduce the total number of training steps and thus save computation resources. Because speech or speaker information is learnt through the encoder of an end-to-end ASR model, we only prune encoder parameters including embedding network, self-attention network and feedforward network in all encoder layers.

Afterwards, we keep the informative sub-network untouched by freezing the unpruned parameters $\theta_{UP}$ to retain the performance on existing speakers, as represented by the light gray connections in Figure~\ref{fig:prune}(c), and only finetune the pruned free parameters $\theta_{P}$ for target speaker adaptation (blue connections in Figure~\ref{fig:prune}(c)). The training objective will be:
\begin{equation}
  \mathcal{J}(\theta_P)=\argmax_{\theta_P}\sum_{j=1}^{L_t}\log P(z_{D_t}^j|y_{D_t}^j;\theta_{UP},\theta_{P})
  \label{asr4}
\end{equation}
where $\theta_{UP}$ is frozen and $\theta_{P}$ is updated. Since the informative sub-network is already capable of performing ASR task very well, we believe further finetuning the free parameters with target speaker data is an added value to the speaker-specific model.

Our method does not change the model architecture, unlike some approaches to attach an additional adapter module \citep{ags}. Besides, we only need to finetune a small number of parameters compared to finetuning the entire model. Fixing the informative sub-network makes our model retain past knowledge with no catastrophic forgetting issue. It also prevents the model from easily overfitting on low-resource target speaker data to some extent.

\section{Experiments}

In this section, we present our experiments using the proposed speaker-aware persistent memory model and the gradual pruning method.

\subsection{Datasets}

We conduct experiments to confirm the effectiveness of the proposed model on the open-source \href{https://www.openslr.org/12}{Librispeech} dataset \citep{librispeech}. LibriSpeech consists of 16kHz read English speech from audiobooks. We use the given train/development/test splits of Librispeech dataset. Test\_clean data is clean and Test\_other data has noise in speech. See Appendix~\ref{sec:librispeechdata} for the statistics of Librispeech dataset used in our experiments.

\begin{table*}
\centering
\begin{tabular}{lcc}
\hline
\textbf{Model}  & \textbf{Test\_clean}  & \textbf{Test\_other}    \\
\hline
End-to-end (E2E) \citep{RWTH_ASR}      & 14.7    & 40.8     \\
E2E with augmented data \citep{literature16}  & 15.1  & -  \\
Local prior matching \citep{Local-prior-matching}   & 14.85    & 39.95     \\
LAS \citep{literature17}                            & 12.9    & 35.5     \\
Self-training \citep{self-training}     & 8.06    & 30.44     \\
Baseline                        & 9.2    & 21.9     \\
\hline
\end{tabular}
\caption{\label{tab:baseline}
WER results of speech recognition models on LibriSpeech 100h.
}
\end{table*}

\subsection{Training Setup}

We use PyTorch and Espnet \citep{espnet} toolkit for our experiments, and we train the model for 100 epochs ($n=100$ in Figure~\ref{fig:prune}(b)). We use the best set of hyperparameters tested by \citet{espnet} for transformer model without further tuning, and we pre-process the data following the Espnet toolkit. The total number of model parameters is 31 million. Input features are generated by 80-dimensional filterbanks with pitch on each frame, with a window size of 25ms shifted every 10ms. The acoustic features are mean and variance normalized. We exclude utterances longer than 3000 frames or 400 characters to keep memory manageable. For joint decoding of CTC and attention, the coefficient is 0.3 and 0.7 for CTC and attention respectively. The convolutional frontend before transformer encoder is two 2D convolutional neural network layers with filter size (3,2) and stride 2, each followed by a ReLU activation. The attention dimension $d_{model}$ is 256, and the feedforward network hidden state dimension $d_{ff}$ is 2048. In the transformer structure, the number of attention heads $h$ is 4, with $d_k=d_q=d_v=64$ for each head, the number of encoder layers $N_e$ is 12, the number of decoder layers $N_d$ is 6, the initial value of learning rate is 5.0, the encoder and decoder dropout rate is 0.1. The input samples are shuffled randomly and trained with batch size 12. We use unigram sub-word algorithm with the vocabulary size capped to be 5000. For i-vector generation, we follow SRE08 recipe in Kaldi \cite{kaldi} toolkit on the training data. I-vectors extracted are of dimension 100. They are then transformed to have the same dimension as speech vectors for concatenation. Our baseline model is competitive compared with other model results from Table~\ref{tab:baseline}.

\subsection{Experimental Results}
\subsubsection{Adaptation for General Speakers}

We first test on the adaptation for general speakers without knowing target speaker profile. Speaker-aware persistent memory model introduced in Section~\ref{sec:speaker_aware} achieves this objective. Here we omit the hyperparameter tuning part, and directly use the best hyperparameters tested by \citet{zyz1}, including the number of speaker i-vectors in the speaker space and the number of layers applied with speaker-aware persistent memory. We randomly sample 64 speaker i-vectors and apply on all the encoder layers in speech transformer. 64 i-vectors were tested to be a good choice to provide diverse speaker information \citep{zyz1}, and applying on all encoder layers helps capture speaker knowledge from both low-level phonetic features and high-level global information. Table~\ref{tab:ivec} shows that our method brings 2.74-6.52\% relative improvement over the baseline, and surpasses \citet{fanzhiyun} who also use speaker i-vectors. Furthermore, here we also compare our model with the first persistent memory model used in ASR \citep{DFSMN-SAN}, in which persistent memory vectors are randomly initialized and meant to capture general knowledge. Different from them, our model is to address the speaker mismatch issue. Our method achieves the best results.

\begin{table}
\centering
\begin{tabular}{lcc}
\hline
\textbf{Model}  & \textbf{Test\_clean}  & \textbf{Test\_other}    \\
\hline
Baseline                        & 9.2    & 21.9     \\
\citet{DFSMN-SAN}               & 8.9    & 21.6     \\
\citet{fanzhiyun}               & 8.9    & 21.4     \\
Ours                            & 8.6    & 21.3     \\
\hline
\end{tabular}
\caption{\label{tab:ivec}
State-of-the-art results of different speaker adaptation algorithms on Librispeech test data.
}
\end{table}

\subsubsection{Adaptation for Target Speaker}

If the target speaker profile is known beforehand, the gradual pruning method discussed in Section~\ref{sec:gradual_pruning} could adapt to the target speaker. Directly finetuning the entire model takes high computation resources by updating all model parameters, and could overfit easily if the amount of target speaker data is limited. We are interested to see the performance of the gradual pruning method especially on low-resource data, as well as how much it alleviates the catastrophic forgetting problem. Therefore, we randomly choose a speaker from the Librispeech Test\_other data as the target speaker, and only select 10 utterances of the target speaker as training data. The remaining utterances of the target speaker are chosen as the test data. We do this four times and report the average performance to see the generalizability of the proposed approach. The average baseline WER of four speakers is 20.5, and is slightly smaller than the average WER of Test\_other speakers, which is 21.9, so further improving the target speaker performance is a bit more challenging. The pruning rate is set as 10\% here. We compare the performance of 1) Finetune: directly finetuning the entire model as Eq.~\ref{asr3}, 2) I-vec: speaker-aware persistent memory method by adding i-vectors, 3) Pruning: gradual pruning, 4) Pruning+I-vec: combining feature adaptation and model adaptation methods proposed.

\begin{figure}[t]
  \centering
  \includegraphics[width=\linewidth]{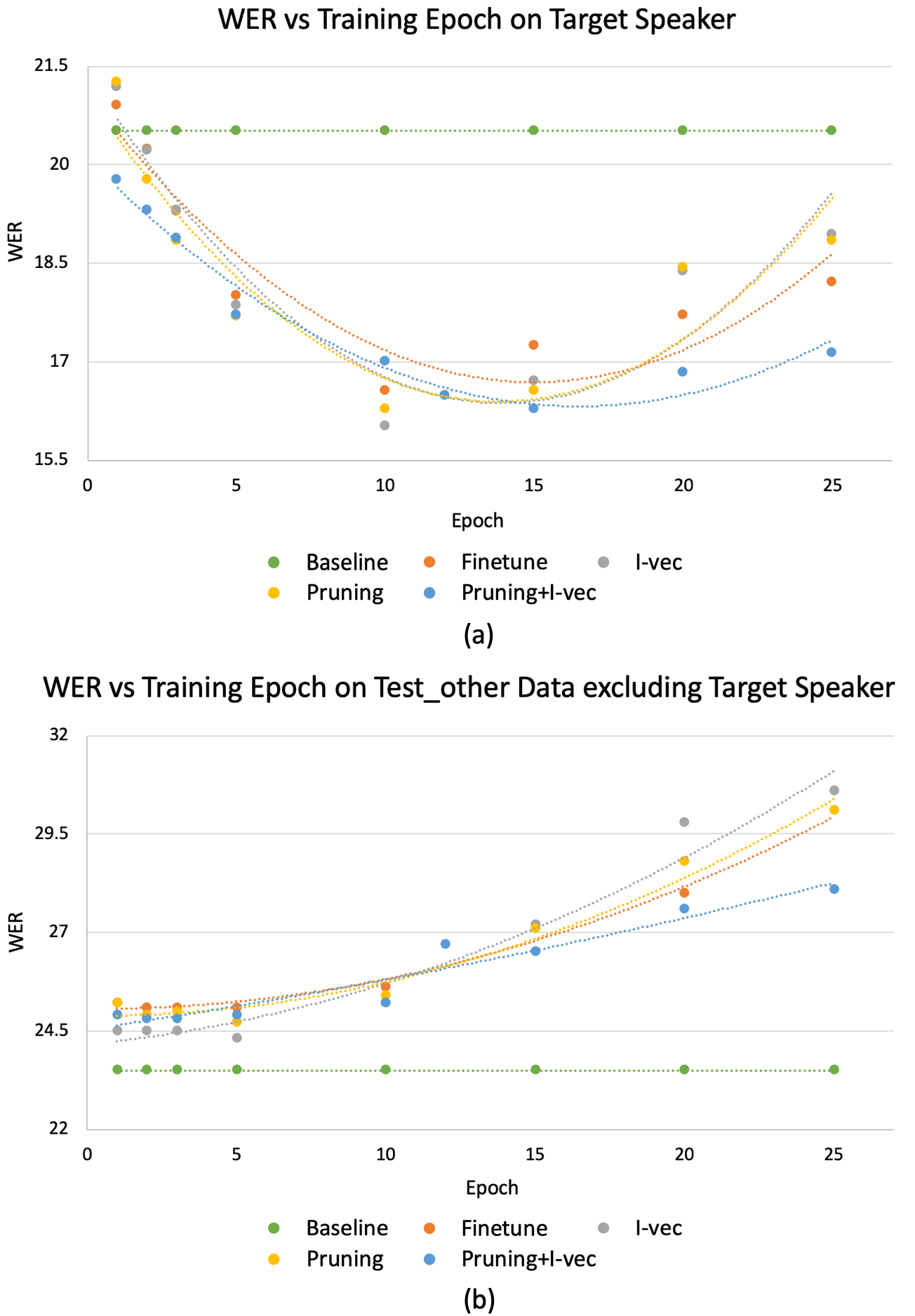}
  \caption{WER results of target speaker (a) and non-target speakers (b). Finetune: directly finetuning the entire model as Eq.~\ref{asr3}. I-vec: speaker-aware persistent memory method proposed in Section~\ref{sec:speaker_aware} by adding i-vectors. Pruning: gradual pruning proposed in Section~\ref{sec:gradual_pruning}. Pruning+I-vec: combining feature adaptation and model adaptation methods proposed. The dotted lines are the second order polynomial trendlines.}
  \label{fig:wer1}
\end{figure}

For results on the target speaker in Figure~\ref{fig:wer1}(a), finetuning works better than the baseline. Adding i-vectors has the highest WER initially and the performance is worse than simply finetuning the trained model after 20 epochs. We believe speaker-aware persistent memory method works better on general speaker adaptation given that the sampled i-vectors form the speaker space to capture any speaker knowledge. It is not designed to adapt to some specific speakers. Using the gradual pruning method alone has lower WER than finetuning at the initial stage, but surprisingly it overfits more than the finetuning method after 20 epochs. More detailed analysis is needed and we leave it to future work. Lastly, we combine the feature adaptation and model adaptation methods, and it achieves our best result. It outperforms the baseline with up to 20.58\% relative WER reduction, and surpasses the finetuning method by up to relative 2.54\%. We see that the feature adaptation method and the model adaptation method we propose complement each other, as the combined model result surpasses each individual one.

We want to analyze the performance of the rest non-target speaker data to see if catastrophic forgetting happens. From Figure~\ref{fig:wer1}(b), all target speaker adapted models perform slightly worse than the baseline, which is expected. Combining feature adaptation and model adaptation could alleviate catastrophic forgetting problem effectively. It generally outperforms finetuning in Figure~\ref{fig:wer1}(b).

\section{Analysis}

In this section, we revisit our approach to reveal more details and explore the effectiveness of the gradual pruning method in combination with the speaker-aware persistent memory model.

\subsection{Pruning Rate}

\begin{figure}[t]
  \centering
  \includegraphics[width=\linewidth]{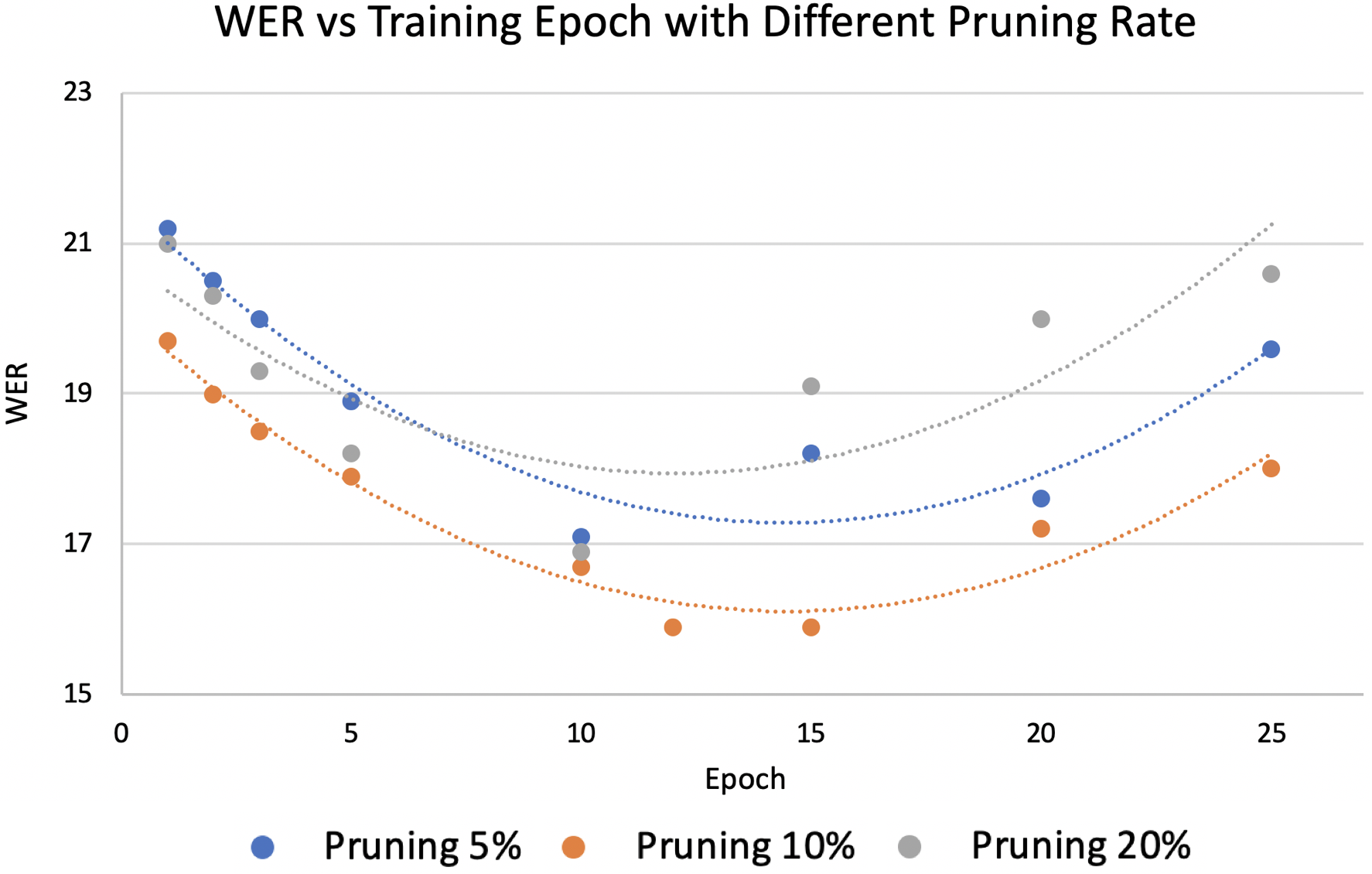}
  \caption{WER results of target speaker with different pruning rates. The dotted lines are the second order polynomial trendlines.}
  \label{fig:pr1}
\end{figure}

We first test different pruning rates on encoder. Results are shown in Figure~\ref{fig:pr1}. Less pruning rate keeps more parameters for the general speaker data, and has less learning capability to target speaker. It is more suitable for simple adaptation tasks. Higher pruning rate generates a more sparse network and is more flexible for speaker adaptation, except that it retains less original model parameters, thus forgets more on the general speaker data. It can be seen from Figure~\ref{fig:pr1} that pruning 10\% of encoder parameters achieves the best result.

\subsection{Gradual Pruning vs One-time Pruning}

We use the gradual pruning method \citep{prune2} to prune to target sparsity for every 10k training steps. One-time pruning at the initial/middle/final stage of the overall training is tested for comparison as well. We train for 100 epochs, and initial/middle/final stage pruning is done at 0/50/100 epoch respectively. Gradual pruning and one-time pruning will reach the same sparsity level after the training. Here we use either gradual or one-time pruning at different stages during training, and show the best results of finetuning for 15 epochs. Table~\ref{tab:gp_onetime} shows that gradual pruning works better than one-time pruning, be it initial, middle or final stage of the training. Compared with one-time pruning, gradual pruning could learn and prune at the same time. In particular, gradual pruning follows the train prune cycle, and is capable of iteratively learning the unpruned parameters after less contributing parameters are pruned. For the one-time pruning, pruning at an earlier stage has the advantage to let the model learn the unpruned parameters based on the pruned ones in the remaining training of the model, but pruning earlier has the risk to prune important parameters since the model is not well learnt yet, vice versa for pruning late. Hence, gradual pruning works the best.

\begin{table}
\small
\centering
\begin{tabular}{lc}
\hline
\textbf{Model}  & \textbf{Target Speaker}      \\
\hline
Baseline                           & 19.9      \\
One-time pruning at initial stage  & 16.2      \\
One-time pruning at middle stage   & 17.6      \\
One-time pruning at final stage    & 16.0      \\
Gradual pruning                    & 15.9      \\
\hline
\end{tabular}
\caption{\label{tab:gp_onetime}
WER results of gradual pruning versus one-time pruning. We train for 100 epochs, and initial/middle/final stage pruning is done at 0/50/100 epoch respectively.
}
\end{table}

\subsection{Extremely Low-resource Adaptation Data}

\begin{table}
\small
\centering
\begin{tabular}{lccc}
\hline
\textbf{Utterance(s)}  & \textbf{1}  & \textbf{5}  & \textbf{10}  \\
\hline
Total No. of Words   & 32          &  104        & 174        \\
Total Duration (s)   & 15.00       &  40.00      & 67.17      \\
\hline
\end{tabular}
\caption{\label{tab:low_resource}
Characteristics of utterances selected as the extremely low-resource adaptation data.
}
\end{table}

\begin{figure}[t]
  \centering
  \includegraphics[width=\linewidth]{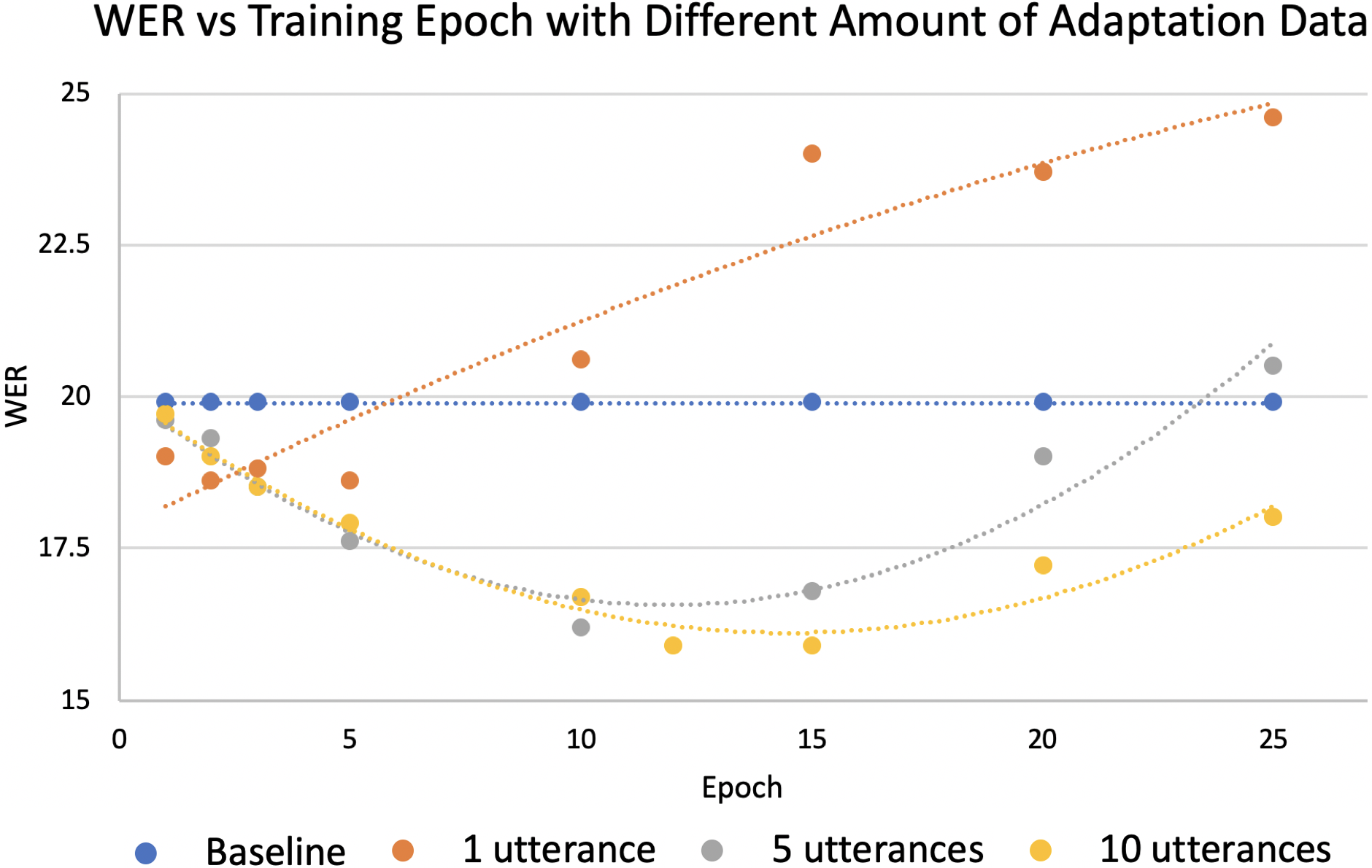}
  \caption{WER results of target speaker with different amount of adaptation data. The dotted lines are the second order polynomial trendlines.}
  \label{fig:pa1}
\end{figure}

Lastly, we would like to see the extremely low-resource adaptation data scenarios. We reduce the amount of adaptation data and compare the performance with the baseline, where no adaptation is performed. The characteristics of the adaptation data selected are listed in Table~\ref{tab:low_resource}. From Figure~\ref{fig:pa1}, when the amount of adaptation data is reduced from 10 utterances to 5 utterances, the results are similar to that of 10 utterances at the initial training stage, and could outperform the baseline by up to relative 18.59\%. With less adaptation data, the model overfits much faster, especially in the case of having only 1 utterance for adaptation. However, even with only 1 utterance, it could surpass the baseline by up to relative 6.53\% with only 5 epochs of training. Therefore, even with extremely low-resource adaptation data such as 1 utterance, our method is effective with fast adaptation.

\section{Conclusion}

In this paper, we have proposed a unified speaker adaptation approach consisting of feature adaptation and model adaptation. Speaker-aware persistent memory model makes use of speaker i-vectors to adapt at the feature level, and we use the gradual pruning approach to retrieve a subset of model parameters for adaptation at the model level. Gradual pruning is found to be better than one-time pruning because gradual pruning could iteratively learn based on pruned parameters. It can alleviate catastrophic forgetting problem as well by retaining a subnetwork whose performance matches the original network. We find that our proposed method is effective in both general speaker adaptation and specific target speaker adaptation. In particular, our method brings relative 2.74-6.52\% WER reduction on general speaker adaptation, and outperforms the baseline with up to 20.58\% relative WER reduction on target speaker adaptation. Even with extremely low-resource adaptation data, our method could bring 6.53\% relative improvement with only a few training epochs. In the future, we are interested in the overfitting issue with low-resource data, as well as multi-speaker adaptation with our method.

\bibliography{anthology,custom}

\begin{thebibliography}{52}
\expandafter\ifx\csname natexlab\endcsname\relax\def\natexlab#1{#1}\fi

\bibitem[{Bérard et~al.(2018)Bérard, Besacier, Kocabiyikoglu, and
  Pietquin}]{literature16}
Alexandre Bérard, Laurent Besacier, Ali~Can Kocabiyikoglu, and Olivier
  Pietquin. 2018.
\newblock \href {https://doi.org/10.1109/ICASSP.2018.8461690} {End-to-end
  automatic speech translation of audiobooks}.
\newblock In \emph{2018 {IEEE} International Conference on Acoustics, Speech
  and Signal Processing ({ICASSP})}. {IEEE}.

\bibitem[{Dehak et~al.(2011{\natexlab{a}})Dehak, A.Torres-Carrasquillo,
  Reynolds, and Dehak}]{ivec1}
Najim Dehak, Pedro A.Torres-Carrasquillo, Douglas Reynolds, and Reda Dehak.
  2011{\natexlab{a}}.
\newblock \href
  {https://www.isca-speech.org/archive/archive_papers/interspeech_2011/i11_0857.pdf}
  {Language recognition via ivectors and dimensionality reduction}.
\newblock In \emph{{INTERSPEECH} 2011 -- 12\textsuperscript{th} Annual
  Conference of the International Speech Communication Association}, pages
  857--860.

\bibitem[{Dehak et~al.(2011{\natexlab{b}})Dehak, Kenny, Dehak, Dumouchel, and
  Ouellet}]{ivec2}
Najim Dehak, Patrick~J Kenny, R{\'{e}}da Dehak, Pierre Dumouchel, and Pierre
  Ouellet. 2011{\natexlab{b}}.
\newblock \href {https://doi.org/10.1109/TASL.2010.2064307} {Front-end factor
  analysis for speaker verification}.
\newblock \emph{{IEEE} Transactions on Audio, Speech, and Language Processing},
  19(4):788--798.

\bibitem[{Ding et~al.(2020)Ding, Guo, Dai, and Du}]{ags}
Fenglin Ding, Wu~Guo, Lirong Dai, and Jun Du. 2020.
\newblock \href {https://doi.org/10.1109/ICASSP40776.2020.9053967}
  {Attention-based gated scaling adaptive acoustic model for {CTC}-based speech
  recognition}.
\newblock In \emph{2020 {IEEE} International Conference on Acoustics, Speech
  and Signal Processing ({ICASSP})}.

\bibitem[{Dong et~al.(2018)Dong, Xu, and Xu}]{Speech-transformer}
Linhao Dong, Shuang Xu, and Bo~Xu. 2018.
\newblock \href {https://doi.org/10.1109/ICASSP.2018.8462506}
  {Speech-transformer: A no-recurrence sequence-to-sequence model for speech
  recognition}.
\newblock In \emph{2018 {IEEE} International Conference on Acoustics, Speech
  and Signal Processing ({ICASSP})}, pages 5884--5888. {IEEE}.

\bibitem[{Fan et~al.(2019)Fan, Li, Zhou, and Xu}]{fanzhiyun}
Zhiyun Fan, Jie Li, Shiyu Zhou, and Bo~Xu. 2019.
\newblock \href {https://doi.org/10.1109/ASRU46091.2019.9003844} {Speaker-aware
  speech-transformer}.
\newblock In \emph{2019 IEEE Automatic Speech Recognition and Understanding
  Workshop (ASRU)}, pages 222--229.

\bibitem[{Frankle and Carbin(2019)}]{prune1}
Jonathan Frankle and Michael Carbin. 2019.
\newblock \href {https://arxiv.org/abs/1803.03635} {The lottery ticket
  hypothesis: Finding sparse, trainable neural networks}.
\newblock In \emph{2019 International Conference on Learning Representations
  ({ICLR})}.

\bibitem[{Furui(1980)}]{hmm2}
S.~Furui. 1980.
\newblock \href {https://doi.org/10.1109/TASSP.1980.1163393} {A training
  procedure for isolated word recognition systems}.
\newblock \emph{{IEEE} Transactions on Audio, Speech, and Language Processing},
  28(2):129--136.

\bibitem[{Gales(1998)}]{speakerspace1}
Mark J.~F. Gales. 1998.
\newblock \href {https://www.isca-speech.org/archive/icslp_1998/i98_0375.html}
  {Cluster adaptive training for speech recognition}.
\newblock \emph{Int. Conf. Speech Language Processing}, 5:1783--1786.

\bibitem[{Gauvain and Lee(1994)}]{hmm3}
J.-L. Gauvain and Chin-Hui Lee. 1994.
\newblock \href {https://doi.org/10.1109/89.279278} {Maximum a posteriori
  estimation for multivariate gaussian mixture observations of markov chains}.
\newblock \emph{{IEEE} Transactions on Audio, Speech, and Language Processing},
  2(2):291--298.

\bibitem[{Graves(2012)}]{literature3}
Alex Graves. 2012.
\newblock \href {https://arxiv.org/abs/1211.3711} {Sequence transduction with
  recurrent neural networks}.
\newblock In \emph{International Conference of Machine Learning (ICML)}, pages
  235--242.

\bibitem[{Gulati et~al.(2020)Gulati, Qin, Chiu, Parmar, Zhang, Yu, Han, Wang,
  Zhang, Wu, and Pang}]{conformer}
Anmol Gulati, James Qin, Chung-Cheng Chiu, Niki Parmar, Yu~Zhang, Jiahui Yu,
  Wei Han, Shibo Wang, Zhengdong Zhang, Yonghui Wu, and Ruoming Pang. 2020.
\newblock \href {https://arxiv.org/abs/2005.08100} {Conformer:
  Convolution-augmented transformer for speech recognition}.
\newblock In \emph{{INTERSPEECH} 2020 -- 21\textsuperscript{st} Annual
  Conference of the International Speech Communication Association}, pages
  5036--5040.

\bibitem[{Han et~al.(2015)Han, Pool, Tran, and Dally}]{prune5}
Song Han, Jeff Pool, John Tran, and William~J. Dally. 2015.
\newblock \href {https://arxiv.org/abs/1506.02626} {Learning both weights and
  connections for efficient neural networks}.
\newblock In \emph{Advances in Neural Information Processing Systems 28: Annual
  Conference on Neural Information Processing Systems}.

\bibitem[{Hsu et~al.(2020)Hsu, Lee, Synnaeve, and
  Hannun}]{Local-prior-matching}
Wei-Ning Hsu, Ann Lee, Gabriel Synnaeve, and Awni Hannun. 2020.
\newblock \href {https://arxiv.org/abs/2002.10336} {Self-supervised speech
  recognition via local prior matching}.
\newblock \emph{arXiv preprint arXiv:2002.10336}.

\bibitem[{Irie et~al.(2019)Irie, Prabhavalkar, Kannan, Bruguier, Rybach, and
  Nguyen}]{literature17}
Kazuki Irie, Rohit Prabhavalkar, Anjuli Kannan, Antoine Bruguier, David Rybach,
  and Patrick Nguyen. 2019.
\newblock \href {https://arxiv.org/abs/1902.01955} {On the choice of modeling
  unit for sequence-to-sequence speech recognition}.
\newblock In \emph{{INTERSPEECH} 2019 -- 20\textsuperscript{th} Annual
  Conference of the International Speech Communication Association}.

\bibitem[{Kahn et~al.(2020)Kahn, Lee, and Hannun}]{self-training}
Jacob Kahn, Ann Lee, and Awni Hannun. 2020.
\newblock \href {https://doi.org/10.1109/ICASSP40776.2020.9054295}
  {Self-training for end-to-end speech recognition}.
\newblock In \emph{2020 {IEEE} International Conference on Acoustics, Speech
  and Signal Processing ({ICASSP})}. {IEEE}.

\bibitem[{Karafiat et~al.(2011)Karafiat, Burget, Matejka, Glembek, and
  Cernocky}]{ivec3}
Martin Karafiat, Lukas Burget, Pavel Matejka, Ondrej Glembek, and Jan Cernocky.
  2011.
\newblock \href {https://doi.org/10.1109/ASRU.2011.6163922} {{iVector}-based
  discriminative adaptation for automatic speech recognition}.
\newblock In \emph{2011 IEEE Automatic Speech Recognition and Understanding
  Workshop (ASRU)}, pages 152--157.

\bibitem[{Keith and Matthias(2005)}]{hmm1}
Johnson Keith and Sjerps Matthias. 2005.
\newblock \href
  {http://linguistics.berkeley.edu/~kjohnson/papers/revised_chapter.pdf}
  {Speaker normalization in speech perception}.
\newblock In \emph{The Handbook of Speech Perception}, pages 363--389. Wiley
  Online Library.

\bibitem[{Kim et~al.(2017)Kim, Song, and Bengio}]{feature11}
Taesup Kim, Inchul Song, and Yoshua Bengio. 2017.
\newblock \href {https://arxiv.org/abs/1707.06065} {Dynamic layer normalization
  for adaptive neural acoustic modeling in speech recognition}.
\newblock In \emph{{INTERSPEECH} 2017 -- 18\textsuperscript{th} Annual
  Conference of the International Speech Communication Association}.

\bibitem[{Kirkpatrick et~al.(2017)Kirkpatrick, Pascanu, Rabinowitz, Veness,
  Desjardins, Rusu, Milan, Quan, Ramalho, Grabska-Barwinska, Hassabis, Clopath,
  Kumaran, and Hadsell}]{cf}
James Kirkpatrick, Razvan Pascanu, Neil Rabinowitz, Joel Veness, Guillaume
  Desjardins, Andrei~A. Rusu, Kieran Milan, John Quan, Tiago Ramalho, Agnieszka
  Grabska-Barwinska, Demis Hassabis, Claudia Clopath, Dharshan Kumaran, and
  Raia Hadsell. 2017.
\newblock \href {https://arxiv.org/abs/1612.00796} {Overcoming catastrophic
  forgetting in neural networks}.
\newblock In \emph{Proceedings of the national academy of sciences}, pages
  3521--3526.

\bibitem[{Kuhn et~al.(2000)Kuhn, Junqua, Nguyen, and Niedzielski}]{hmm4}
Roland Kuhn, Jean-Claude Junqua, Patrick Nguyen, and Nancy Niedzielski. 2000.
\newblock \href {https://doi.org/10.1109/89.876308} {Rapid speaker adaptation
  in eigenvoice space}.
\newblock \emph{{IEEE} Transactions on Audio, Speech, and Language Processing},
  8(6):695--707.

\bibitem[{Li et~al.(2017)Li, Kadav, Durdanovic, Samet, and Graf}]{prune4}
Hao Li, Asim Kadav, Igor Durdanovic, Hanan Samet, and Hans~Peter Graf. 2017.
\newblock \href {https://arxiv.org/abs/1608.08710} {Pruning filters for
  efficient convnets}.
\newblock In \emph{2017 International Conference on Learning Representations
  ({ICLR})}.

\bibitem[{Liang et~al.(2021)Liang, Zhao, Wang, Qiu, and Li}]{prune-tune}
Jianze Liang, Chengqi Zhao, Mingxuan Wang, Xipeng Qiu, and Lei Li. 2021.
\newblock \href {https://arxiv.org/abs/2012.10586} {Finding sparse structures
  for domain specific neural machine translation}.
\newblock In \emph{The Thirty-Fifth AAAI Conference on Artificial Intelligence,
  AAAI 2021}.

\bibitem[{Liao(2013)}]{l2}
Hank Liao. 2013.
\newblock \href {https://doi.org/10.1109/ICASSP.2013.6639212} {Speaker
  adaptation of context dependent deep neural networks}.
\newblock In \emph{2013 {IEEE} International Conference on Acoustics, Speech
  and Signal Processing ({ICASSP})}, pages 7947--7951.

\bibitem[{Liu et~al.(2019)Liu, Sun, Zhou, Huang, and Darrell}]{prune3}
Zhuang Liu, Mingjie Sun, Tinghui Zhou, Gao Huang, and Trevor Darrell. 2019.
\newblock \href {https://arxiv.org/abs/1810.05270} {Rethinking the value of
  network pruning}.
\newblock In \emph{2019 International Conference on Learning Representations
  ({ICLR})}.

\bibitem[{Lüscher et~al.(2019)Lüscher, Beck, Irie, Kitza, Michel, Zeyer,
  Schlüter, and Ney}]{RWTH_ASR}
Christoph Lüscher, Eugen Beck, Kazuki Irie, Markus Kitza, Wilfried Michel,
  Albert Zeyer, Ralf Schlüter, and Hermann Ney. 2019.
\newblock \href {https://doi.org/10.21437/Interspeech.2019-1780} {{RWTH} asr
  systems for librispeech: Hybrid vs attention}.
\newblock In \emph{{INTERSPEECH} 2019 -- 20\textsuperscript{th} Annual
  Conference of the International Speech Communication Association}, pages
  231--235.

\bibitem[{McCloskey and Cohen(1989)}]{catastrophic}
Michael McCloskey and Neal~J. Cohen. 1989.
\newblock \href {https://doi.org/10.1016/S0079-7421(08)60536-8} {Catastrophic
  interference in connectionist networks: The sequential learning problem}.
\newblock 24:109--165.

\bibitem[{Meng et~al.(2019)Meng, Li, and Gong}]{mtl}
Zhong Meng, Jinyu Li, and Yifan Gong. 2019.
\newblock \href {https://arxiv.org/abs/1904.12407} {Adversarial speaker
  adaptation}.
\newblock In \emph{2019 {IEEE} International Conference on Acoustics, Speech
  and Signal Processing ({ICASSP})}, pages 5721--5725.

\bibitem[{Miao et~al.(2015)Miao, Gowayyed, and Metze}]{literature1}
Yajie Miao, Mohammad Gowayyed, and Florian Metze. 2015.
\newblock \href {https://doi.org/10.1109/ASRU.2015.7404790} {{EESEN}:
  End-to-end speech recognition using deep rnn models and wfst-based decoding}.
\newblock In \emph{2015 IEEE Automatic Speech Recognition and Understanding
  Workshop (ASRU)}, pages 167--174.

\bibitem[{Ochiai et~al.(2018)Ochiai, Watanabe, Katagiri, Hori, and
  Hershey}]{feature4}
Tsubasa Ochiai, Shinji Watanabe, Shigeru Katagiri, Takaaki Hori, and John
  Hershey. 2018.
\newblock \href {https://doi.org/10.1109/ICASSP.2018.8462161} {Speaker
  adaptation for multichannel end-to-end speech recognition}.
\newblock In \emph{2018 {IEEE} International Conference on Acoustics, Speech
  and Signal Processing ({ICASSP})}, pages 6707--6711. {IEEE}.

\bibitem[{Pan et~al.(2018)Pan, Liu, Wan, Du, Liu, and Ye}]{feature8}
Jia Pan, Diyuan Liu, Genshun Wan, Jun Du, Qingfeng Liu, and Zhongfu Ye. 2018.
\newblock \href {https://doi.org/10.23919/APSIPA.2018.8659609} {Online speaker
  adaptation for {LVCSR} based on attention mechanism}.
\newblock \emph{2018 Asia-Pacific Signal and Information Processing Association
  Annual Summit and Conference (APSIPA ASC)}, pages 183--186.

\bibitem[{Panayotov et~al.(2015)Panayotov, Chen, Povey, and
  Khudanpur}]{librispeech}
Vassil Panayotov, Guoguo Chen, Daniel Povey, and Sanjeev Khudanpur. 2015.
\newblock Librispeech: An {ASR} corpus based on public domain audio books.
\newblock In \emph{2015 {IEEE} International Conference on Acoustics, Speech
  and Signal Processing ({ICASSP})}. {IEEE}.

\bibitem[{Povey et~al.(2011)Povey, Ghoshal, Boulianne, Burget, Glembek, Goel,
  Hannemann, Motlicek, Qian, Schwarz, Silovsky, Stemmer, and Vesely}]{kaldi}
Daniel Povey, Arnab Ghoshal, Gilles Boulianne, Lukas Burget, Ondrej Glembek,
  Nagendra Goel, Mirko Hannemann, Petr Motlicek, Yanmin Qian, Petr Schwarz, Jan
  Silovsky, Georg Stemmer, and Karel Vesely. 2011.
\newblock \href {https://www.danielpovey.com/files/2011_asru_kaldi.pdf} {The
  kaldi speech recognition toolkit}.
\newblock In \emph{2011 IEEE Automatic Speech Recognition and Understanding
  Workshop (ASRU)}.

\bibitem[{Samarakoon and Sim(2016{\natexlab{a}})}]{model3}
Lahiru Samarakoon and Khe~Chai Sim. 2016{\natexlab{a}}.
\newblock \href {https://doi.org/10.1109/TASLP.2016.2601146} {Factorized hidden
  layer adaptation for deep neural network based acoustic modeling}.
\newblock \emph{{IEEE}/{ACM} Transactions on Audio, Speech, and Language
  Processing}, 24(12):2241--2250.

\bibitem[{Samarakoon and Sim(2016{\natexlab{b}})}]{lhuc2}
Lahiru Samarakoon and Khe~Chai Sim. 2016{\natexlab{b}}.
\newblock \href {https://doi.org/10.21437/Interspeech.2016-1249} {Subspace
  {LHUC} for fast adaptation of deep neural network acoustic models}.
\newblock In \emph{{INTERSPEECH} 2016 -- 17\textsuperscript{th} Annual
  Conference of the International Speech Communication Association}, pages
  1593--1597.

\bibitem[{Saon et~al.(2013)Saon, Soltau, Nahamoo, and Picheny}]{feature5}
George Saon, Hagen Soltau, David Nahamoo, and Michael Picheny. 2013.
\newblock \href {https://doi.org/10.1109/ASRU.2013.6707705} {Speaker adaptation
  of neural network acoustic models using i-vectors}.
\newblock In \emph{2013 IEEE Automatic Speech Recognition and Understanding
  Workshop (ASRU)}.

\bibitem[{Seide et~al.(2011)Seide, Li, Chen, and Yu}]{feature2}
Frank Seide, Gang Li, Xie Chen, and Dong Yu. 2011.
\newblock \href {https://doi.org/10.1109/ASRU.2011.6163899} {Feature
  engineering in context-dependent deep neural networks for conversational
  speech transcription}.
\newblock In \emph{2011 IEEE Automatic Speech Recognition and Understanding
  Workshop (ASRU)}.

\bibitem[{Senior and Lopez-Moreno(2014)}]{feature7}
Andrew Senior and Ignacio Lopez-Moreno. 2014.
\newblock \href {https://doi.org/10.1109/ICASSP.2014.6853591} {Improving {DNN}
  speaker independence with i-vector inputs}.
\newblock In \emph{2014 {IEEE} International Conference on Acoustics, Speech
  and Signal Processing ({ICASSP})}.

\bibitem[{Siniscalchi et~al.(2013)Siniscalchi, Li, and Lee}]{model2}
Sabato~Marco Siniscalchi, Jinyu Li, and Chin-Hui Lee. 2013.
\newblock \href {https://doi.org/10.1109/TASL.2013.2270370} {Hermitian
  polynomial for speaker adaptation of connectionist speech recognition
  systems}.
\newblock \emph{{IEEE} Transactions on Audio, Speech, and Language Processing},
  21(10):2152--2161.

\bibitem[{Swietojanski et~al.(2016)Swietojanski, Li, and Renals}]{lhuc1}
Pawel Swietojanski, Jinyu Li, and Steve Renals. 2016.
\newblock \href {https://doi.org/10.1109/SLT.2014.7078569} {Learning hidden
  unit contributions for unsupervised acoustic model adaptation}.
\newblock \emph{IEEE/ACM Transactions on Audio, Speech and Language
  Processing}, 24(8):1450--1463.

\bibitem[{Tomashenko and Estève(2018)}]{feature3}
Natalia Tomashenko and Yannick Estève. 2018.
\newblock \href {https://www.aclweb.org/anthology/L18-1500} {Evaluation of
  feature-space speaker adaptation for end-to-end acoustic models}.
\newblock In \emph{Proceedings of the Eleventh International Conference on
  Language Resources and Evaluation (LREC 2018)}.

\bibitem[{Vaswani et~al.(2017)Vaswani, Shazeer, Parmar, Uszkoreit, Jones,
  Gomez, Kaiser, and Polosukhin}]{transformer}
Ashish Vaswani, Noam Shazeer, Niki Parmar, Jakob Uszkoreit, Llion Jones,
  Aidan~N. Gomez, Lukasz Kaiser, and Illia Polosukhin. 2017.
\newblock \href {https://arxiv.org/abs/1706.03762} {Attention is all you need}.
\newblock In \emph{Advances in Neural Information Processing Systems}, pages
  5998--6008.

\bibitem[{Vesely et~al.(2016)Vesely, Watanabe, Zmolikova, Karafiat, Burget, and
  Cernocky}]{feature10}
Karel Vesely, Shinji Watanabe, Katerina Zmolikova, Martin Karafiat, Lukas
  Burget, and Jan~Honza Cernocky. 2016.
\newblock \href {https://doi.org/10.1109/ICASSP.2016.7472692} {Sequence
  summarizing neural network for speaker adaptation}.
\newblock In \emph{2016 {IEEE} International Conference on Acoustics, Speech
  and Signal Processing ({ICASSP})}, pages 5315--5319. {IEEE}.

\bibitem[{Watanabe et~al.(2018)Watanabe, Hori, Karita, Hayashi, Nishitoba,
  Unno, Soplin, Heymann, Wiesner, Chen, Renduchintala, and Ochiai}]{espnet}
Shinji Watanabe, Takaaki Hori, Shigeki Karita, Tomoki Hayashi, Jiro Nishitoba,
  Yuya Unno, Nelson Enrique~Yalta Soplin, Jahn Heymann, Matthew Wiesner, Nanxin
  Chen, Adithya Renduchintala, and Tsubasa Ochiai. 2018.
\newblock \href {https://arxiv.org/abs/1804.00015} {Espnet: End-to-end speech
  processing toolkit}.
\newblock In \emph{{INTERSPEECH} 2018 -- 19\textsuperscript{th} Annual
  Conference of the International Speech Communication Association}, pages
  2207--2211.

\bibitem[{Xie et~al.(2019)Xie, Liu, Lee, Hu, and Wang}]{lhuc3}
Xurong Xie, Xunying Liu, Tan Lee, Shoukang Hu, and Lan Wang. 2019.
\newblock \href {https://doi.org/10.1109/ICASSP.2019.8682667} {{BLHUC}:
  Bayesian learning of hidden unit contributions for deep neural network
  speaker adaptation}.
\newblock In \emph{2019 {IEEE} International Conference on Acoustics, Speech
  and Signal Processing ({ICASSP})}, pages 5711--5715.

\bibitem[{Yao et~al.(2012)Yao, Yu, Seide, Su, Deng, and Gong}]{model1}
Kaisheng Yao, Dong Yu, Frank Seide, Hang Su, Li~Deng, and Yifan Gong. 2012.
\newblock \href {https://doi.org/10.1109/SLT.2012.6424251} {Adaptation of
  context-dependent deep neural networks for automatic speech recognition}.
\newblock In \emph{2012 {IEEE} Spoken Language Technology Workshop ({SLT})}.
  {IEEE}.

\bibitem[{You et~al.(2019)You, Su, Chen, Weng, and Yu}]{DFSMN-SAN}
Zhao You, Dan Su, Jie Chen, Chao Weng, and Dong Yu. 2019.
\newblock \href {https://arxiv.org/pdf/1910.13282.pdf} {D{FSMN}-{SAN} with
  persistent memory model for automatic speech recognition}.
\newblock \emph{arXiv preprint arXiv:1910.13282}.

\bibitem[{Yu et~al.(2013)Yu, Yao, Su, Li, and Seide}]{kld}
Dong Yu, Kaisheng Yao, Hang Su, Gang Li, and Frank Seide. 2013.
\newblock \href {https://doi.org/10.1109/ICASSP.2013.6639201} {{KL}-divergence
  regularized deep neural network adaptation for improved large vocabulary
  speech recognition}.
\newblock In \emph{2013 {IEEE} International Conference on Acoustics, Speech
  and Signal Processing ({ICASSP})}, pages 7893--7897.

\bibitem[{Yu and Gales(2006)}]{speakerspace2}
Kai Yu and Mark J.~F. Gales. 2006.
\newblock \href {https://doi.org/10.1109/TSA.2005.858555} {Discriminative
  cluster adaptive training}.
\newblock \emph{{IEEE} Transactions on Audio, Speech, and Language Processing},
  14(5):1694--1703.

\bibitem[{Zhang et~al.(2017)Zhang, Chan, and Jaitly}]{literature2}
Yu~Zhang, William Chan, and Navdeep Jaitly. 2017.
\newblock \href {https://doi.org/10.1109/ICASSP.2017.7953077} {Very deep
  convolutional networks for end-to-end speech recognition}.
\newblock In \emph{2017 {IEEE} International Conference on Acoustics, Speech
  and Signal Processing ({ICASSP})}, pages 4845--4849. {IEEE}.

\bibitem[{Zhao et~al.(2020)Zhao, Ni, Leung, Joty, Chng, and Ma}]{zyz1}
Yingzhu Zhao, Chongjia Ni, Cheung-Chi Leung, Shafiq Joty, Eng~Siong Chng, and
  Bin Ma. 2020.
\newblock \href
  {https://www.isca-speech.org/archive/Interspeech_2020/pdfs/1281.pdf} {Speech
  transformer with speaker aware persistent memory}.
\newblock In \emph{{INTERSPEECH} 2020 -- 21\textsuperscript{st} Annual
  Conference of the International Speech Communication Association}, pages
  1261--1265.

\bibitem[{Zhu and Gupta(2018)}]{prune2}
Michael Zhu and Suyog Gupta. 2018.
\newblock \href {https://arxiv.org/abs/1710.01878} {To prune, or not to prune:
  exploring the efficacy of pruning for model compression}.
\newblock In \emph{2018 International Conference on Learning Representations
  ({ICLR})}.

\end{thebibliography}
\bibliographystyle{acl_natbib}

\appendix

\section{Appendix}
\label{sec:appendix}

\subsection{Details of Librispeech dataset}
\label{sec:librispeechdata}

We use open-source Librispeech dataset for all our experiments, which is downloadable from \url{https://www.openslr.org/12}. Table~\ref{tab:dataset} shows the statistics of the dataset.

\subsection{Average Runtime}

In Table~\ref{tab:runtime}, we list the average runtime using one V100 GPU of 1) Baseline, 2) Finetune: directly finetuning the trained baseline model, 3) I-vec: speaker-aware persistent memory method by adding i-vectors, 4) Pruning: gradual pruning, 5) Pruning+I-vec: combining feature adaptation and model adaptation methods proposed. During training, all models are trained with the given 100h Librispeech training data, while during adaptation, all models are trained with 10 utterances of adaptation data, except for the baseline where no adaptation is performed.

\subsection{Evaluation Metrics}

We evaluate model performance by word error rate (WER), which can be computed as following:
\begin{equation}
  WER = \frac{S+D+I}{N_r} = \frac{S+D+I}{S+D+C}
  \label{wer}
\end{equation}
where $S$ is the number of substitutions, $D$ is number of deletions, $I$ is the number of insertions, $N_r$ is number of words in the reference ($N_r = S+D+C$), $C$ is the number of correct words.

\subsection{Computing Infrastructure}

We conduct our experiments on NVIDIA V100 GPU and Intel(R) Xeon(R) Platinum 8163 32-core CPU @ 2.50GHz.

\begin{table}
\centering
\begin{tabular}{ll}
\hline
\textbf{LibriSpeech} &  \\
\hline
Training set & 100h (251 speakers) \\
Dev\_clean set  & 5.4h (20 males, 20 females) \\
Dev\_other set  & 5.3h (17 males, 16 females) \\
Test\_clean set  & 5.4h (20 males, 20 females) \\
Test\_other set  & 5.1h (16 males, 17 females) \\
\hline
\end{tabular}
\caption{\label{tab:dataset}
Statistics of Librispeech dataset used for experiments.
}
\end{table}

\begin{table}
\centering
\begin{tabular}{lcc}
\hline
\textbf{Algorithm} & \textbf{Training} & \textbf{Adaptation} \\
\hline
Baseline       & 2d13h &  - \\
Finetune       & 2d13h &  2min \\
I-vec          & 2d12h &  2min \\
Pruning        & 2d5h  &  2min \\
Pruning+I-vec  & 2d5h  &  2min \\
\hline
\end{tabular}
\caption{\label{tab:runtime}
Average runtime.
}
\end{table}

\end{document}